\documentclass{article}
\usepackage{amssymb}

\usepackage{amsmath}

\newtheorem{corollary}{Corollary}

\newtheorem{proposition}{Proposition}

\numberwithin{equation}{section}
\input{tcilatex}

\begin{document}

\title{Smooth Solutions of the Three Dimensional Navier-Stokes Problem}
\author{Assane Lo \\
%EndAName
King Fahd University of Petroleum and Minerals}
\maketitle

\begin{abstract}
The aim of this paper is to solve the three dimensional Navier-Stokes
problem with conservative source term when the initial conditions are
divergence and curl free.

We use convolution methods with basic vector calculus to construct ``well
behaved'' smooth solutions of the initial boundary value problem for the
system of Navier-Stokes.
\end{abstract}

\section{Introduction}

The Navier-Stokes equations are the equations that describe the motion of
usual fluids like water, air, oil. They appear in the study of many
important phenomena. while the physical model leading to the Navier-Stokes
equations is simple, the situation is quite different from the mathematical
point of view. In particular, because of their nonlinearity, the
mathematical study of these equations appeared difficult and has been open
for many years.

In this paper, we study convolution techniques for solving the 3-dimensional
Navier-Stokes equation when the source term is given by a conservative field.

\section{Constructing Smooth Solutions}

Consider the initial boundary value problem for the system of Navier-Stokes%
\begin{equation}
\left\{ 
\begin{tabular}{l}
$\frac{\partial \mathbf{u}}{\partial t}-\nu \mathbf{\Delta u}+\left( \mathbf{%
u\cdot \nabla }\right) \mathbf{u}+\mathbf{\nabla }p=\mathbf{f}$ \\ 
$\mathbf{divu}=0,$%
\end{tabular}%
\right.
\end{equation}%
with initial conditions 
\begin{equation}
\mathbf{u}(x,0)=\mathbf{u}^{0}(x),\;\;\;\;\;\;\;\;x\in \mathbb{R}^{3}.
\end{equation}%
Here, $\mathbf{u}^{0}(x)$ is a given, $C^{\infty }$ divergence-free vector
field on $\mathbb{R}^{3}.$

This system describes the velocity $\mathbf{u}$ and pressure $p$ of a
viscous incompressible fluid under the influence of an external force $%
\mathbf{f}$. The viscosity $\nu $ is assumed to be constant.

The system of equations (1.1) is to be solved for an unknown velocity vector 
\begin{equation*}
\mathbf{u}(x,t)=(u^{i}(x,t))\in \mathbb{R}^{3}
\end{equation*}%
and pressure $p(x,t)\in \mathbb{R}$, defined for position $x$ $\in $ $%
\mathbb{R}^{3}$ and time $t\geq 0.$

\begin{proposition}
Let $\mathbf{u}^{0}(x)\;$be$\;$a smooth vector field in $\mathbb{R}^{3}$
satisfying 
\begin{equation*}
\mathbf{div\mathbf{u}}^{0}\mathbf{(}x\mathbf{)=curl\mathbf{u}}^{0}\mathbf{(}x%
\mathbf{)}=0
\end{equation*}%
If $\mathbf{f}$ is a conservative force, then there exist $\mathbf{u(}x,t%
\mathbf{),\;}p(x,t)\in C^{\infty }(\mathbb{R}^{3}\times \mathbb{R})$ with $%
\mathbf{u(}x,0\mathbf{)=u}^{0}\mathbf{(}x\mathbf{)}$ solutions of the
Navier-Stokes equation%
\begin{equation*}
\left\{ 
\begin{tabular}{l}
$\frac{\partial \mathbf{u}}{\partial t}-\nu \mathbf{\Delta u}+\left( \mathbf{%
u\cdot \nabla }\right) \mathbf{u}+\mathbf{\nabla }p=\mathbf{f}$ \\ 
$\mathbf{divu}=0.$%
\end{tabular}%
\right.
\end{equation*}
\end{proposition}

\textbf{Proof} Without loss of generality, we may assume that$\ \mathbf{f}=%
\mathbf{0.}$ First observe that 
\begin{equation*}
\left( \mathbf{u\cdot \nabla }\right) \mathbf{u=\nabla }\left( \frac{1}{2}%
\left| \mathbf{u}\right| ^{2}\right) -\mathbf{u\times curlu.}
\end{equation*}%
Equation $(1.1)$ becomes 
\begin{equation*}
\left\{ 
\begin{tabular}{l}
$\frac{\partial \mathbf{u}}{\partial t}-\nu \mathbf{\Delta u}+\mathbf{\nabla 
}\left( \frac{1}{2}\left| \mathbf{u}\right| ^{2}\right) -\mathbf{u\times
curlu}+\mathbf{\nabla }p=\mathbf{0}$ \\ 
$\mathbf{divu}=0.$%
\end{tabular}%
\right.
\end{equation*}%
Let $h$ be a smooth function with compact support satisfying%
\begin{equation*}
\int_{\mathbb{R}^{3}}h(x)dx=1
\end{equation*}%
Let 
\begin{equation*}
\mathbf{u}(x,t)=\int_{\mathbb{R}^{3}}\mathbf{u}^{0}(x-ty)h(y)dy.
\end{equation*}%
$\mathbf{u}(x,t)$ is $C^{\infty }$ in $\left\{ \left( x,t\right) \in \mathbb{%
R}^{3+1}:\;t\neq 0\right\} ,$ satisfies $\mathbf{u}(x,0)=\mathbf{u}^{0}%
\mathbf{(}x\mathbf{)}.$ Moreover, we have $\mathbf{divu}=0$ and $\mathbf{%
curlu}=\mathbf{0.}$ This implies that%
\begin{equation*}
\mathbf{curl}\left[ \frac{\partial \mathbf{u}}{\partial t}-\nu \mathbf{%
\Delta u}+\mathbf{\nabla }\left( \frac{1}{2}\left| \mathbf{u}\right|
^{2}\right) -\mathbf{u\times curlu}\right] =\mathbf{0.}
\end{equation*}%
Hence, the vector field 
\begin{equation*}
\frac{\partial \mathbf{u}}{\partial t}-\nu \mathbf{\Delta u}+\mathbf{\nabla }%
\left( \frac{1}{2}\left| \mathbf{u}\right| ^{2}\right) -\mathbf{u\times curlu%
}
\end{equation*}%
is conservative for all $t.$ Therefore equation $(1.1)$ is satisfied for
some $p(x,t)\in C^{\infty }(\mathbb{R}^{3+1}).\;\;\;\;\;\;\;\;\;\;\;\;\;\;%
\blacksquare $

In [3] a class of radial measures $\mu $ on $\mathbb{R}^{n}$ is defined so
that integrable harmonic functions $g$ on $\mathbb{R}^{n}$ may be
characterize as solutions of convolution equations in $\mathbb{R}^{n}$. In
particular, Natan Y. B. and Weit Y showed that if 
\begin{equation*}
\varphi (x)=c_{n}e^{-2\pi \left| x\right| }
\end{equation*}%
where $c_{n}=\frac{\pi ^{(n+1)/2}}{\Gamma (\frac{n+1}{2})}$ then every
solution of the equation $g\ast h=g$ in $L^{1}(\mathbb{R}^{n},e^{-2\pi
\left| x\right| }dx)$ is harmonic if and only if $n<9.$

\begin{corollary}
Let%
\begin{equation*}
\mathbf{u}^{0}(x)=\mathbf{\nabla }g(x)\;\;\;\;x\in \mathbb{R}^{3}
\end{equation*}%
where $g$ is a solution of the equation $g\ast h=g$ in $L^{1}(\mathbb{R}%
^{3},e^{-2\pi \left| x\right| }dx).$ If $\mathbf{f}$ is a conservative
force, then there exist $\mathbf{u(}x,t\mathbf{),\;}p(x,t)\in C^{\infty }(%
\mathbb{R}^{3}\times \mathbb{R})$ with $\mathbf{u(}x,0\mathbf{)=u}^{0}%
\mathbf{(}x\mathbf{)}$ solutions of the Navier-Stokes equation%
\begin{equation*}
\left\{ 
\begin{tabular}{l}
$\frac{\partial \mathbf{u}}{\partial t}-\nu \mathbf{\Delta u}+\left( \mathbf{%
u\cdot \nabla }\right) \mathbf{u}+\mathbf{\nabla }p=\mathbf{f}$ \\ 
$\mathbf{divu}=0.$%
\end{tabular}%
\right.
\end{equation*}
\end{corollary}

\section{Towards a Well-behaved Solution}

In this section, we propose to construct smooth solutions $\mathbf{u}%
(x,y,z,t)$ in $\mathbb{R}^{3+1}$ with a suitable growth condition.

\begin{proposition}
There exist smooth vector fields $\mathbf{u}(x,y,z,t)\in C^{\infty }(\mathbb{%
R}^{3+1},\mathbb{R}^{3}),$ and a smooth functions $p(x,t)\in C^{\infty }(%
\mathbb{R}^{3+1})$ solutions of equation (1.1) with $\mathbf{u}(x,y,z,t)$
periodic in $x$ and $y$, and whose behavior in $z$ may be controlled by a
smooth function $\Psi :\mathbb{R\rightarrow R}$ satisfying $\Psi ^{\prime
\prime }=\Psi $.
\end{proposition}

\textbf{Proof.} Let $\alpha ,\;\beta ,\;$and $\zeta $ be nonzero real
numbers such that 
\begin{equation*}
\zeta ^{2}=\alpha ^{2}+\beta ^{2}.
\end{equation*}%
Let $\varphi $ be an even smooth function with compact support in $\mathbb{R}%
.$ Let $g$ be the function defined by 
\begin{equation*}
g(x,y,z)=\left( \int_{\mathbb{R}}\varphi (z-r)\Psi \left( \zeta r\right)
dr\right) \cos (\alpha x+\beta y).
\end{equation*}%
$g$ is a harmonic function in $\mathbb{R}^{3}$ that is periodic in $x,$ and $%
y$.

Now define 
\begin{equation*}
\mathbf{u}^{0}(\mathbf{x})=\nabla g(\mathbf{x})
\end{equation*}%
where $\mathbf{x}=(x,y,z)\in \mathbb{R}^{3},$ and 
\begin{equation*}
\mathbf{u}(\mathbf{x,}t)=\int_{\mathbb{R}^{3}}\mathbf{u}^{0}(\mathbf{x}-t%
\mathbf{y})h(\mathbf{y})d\mathbf{y.\;\;\;}t\neq 0.
\end{equation*}%
$h$ is as in proposition 1 a smooth function with compact support satisfying 
$\int_{\mathbb{R}^{3}}h(x)dx=1.$

The proof of proposition 1 shows that $\mathbf{u}(\mathbf{x,}t)$ and $p(x,t)$
will solve equation (1.1) for some smooth function $p(x,t)$ in $\mathbb{R}%
^{3+1}.$

Moreover $\mathbf{u}(\mathbf{x,}t)$ satisfies the desired requirement. That
is, $\mathbf{u}(x,y,z,t)$ is periodic in $x$ and $y$, and whose behavior in $%
z$ depends on the growth of the function $\Psi $.$\blacksquare $

\bigskip 

\textbf{Conlusion. }The results above guarantee the existence of smooth
solutions to the three dimensional Navier-Stoke problem. A fundamental
problem in analysis is to decide whether a smooth, physically reasonable
solutions in the sense of [1] exist for the Navier--Stokes equations. We
believed that we have given a partial answer to the question by constructing
solutions that behave well in the two variables with initial conditions that
are both divergence and curl free.

\end{document}